\documentclass[a4paper]{cas-sc}

\usepackage{cite}
\usepackage{amsmath,amssymb,amsfonts}
\usepackage{algorithmic}
\usepackage{graphicx}
\usepackage{textcomp}
\usepackage[ruled,linesnumbered]{algorithm2e}
\usepackage{diagbox} 
\usepackage{makecell} 
\usepackage{booktabs}
\usepackage{subfigure}  
\usepackage{longtable}
\usepackage{threeparttable}
\usepackage[numbers]{natbib}

\def\tsc#1{\csdef{#1}{\textsc{\lowercase{#1}}\xspace}}
\tsc{WGM}
\tsc{QE}
\tsc{EP}
\tsc{PMS}
\tsc{BEC}
\tsc{DE}

\begin{document}
\let\WriteBookmarks\relax
\def\floatpagepagefraction{1}
\def\textpagefraction{.001}
\shorttitle{DroidRL}
\shortauthors{Wu et~al.}

\title [mode = title]{DroidRL: Feature Selection for Android Malware Detection with Reinforcement Learning}                      
\tnotemark[1]

\author[1]{Yinwei Wu}
\author[1]{Meijin Li}
\author[3]{Qi Zeng}
\author[3]{Tao Yang}
\author[3]{Junfeng Wang}
\author[2]{Zhiyang Fang}
\cormark[1]
\ead{fangzhiyang@scu.edu.cn}
\author[4]{Luyu Cheng}

\address[1]{College of Software Engineering, Sichuan University, Chengdu, China}
\address[2]{School of Cyber Science and Engineering, Sichuan University, Chengdu , China}
\address[3]{College of Computer Science, Sichuan University, Chengdu, China}
\address[4]{School of Business, Sichuan University, Chengdu, China}

\begin{abstract}
Due to the completely open-source nature of Android, the exploitable vulnerability of malware attacks is increasing. Machine learning, leading to a great evolution in Android malware detection in recent years, is typically applied in the classification phase. Since the correlation between features is ignored in some traditional ranking-based feature selection algorithms, applying wrapper-based feature selection models is a topic worth investigating. Though considering the correlation between features, wrapper-based approaches are time-consuming for exploring all possible valid feature subsets when processing a large number of Android features. To reduce the computational expense of wrapper-based feature selection, a framework named DroidRL is proposed. The framework deploys DDQN algorithm to obtain a subset of features which can be used for effective malware classification. To select a valid subset of features over a larger range, the exploration-exploitation policy is applied in the model training phase. The recurrent neural network (RNN) is used as the decision network of DDQN to give the framework the ability to sequentially select features. Word embedding is applied for feature representation to enhance the framework’s ability to find the semantic relevance of features. The framework’s feature selection exhibits high performance without any human intervention and can be ported to other feature selection tasks with minor changes. The experiment results show a significant effect when using the Random Forest as DroidRL’s classifier, which reaches 95.6\% accuracy with only 24 features selected.

\end{abstract}


\maketitle

\section{INTRODUCTION}

Android is the fastest-growing computing platform on the mobile terminal. In 2021, there were 1.39 billion smartphones manufactured worldwide, and Android dominated the global market by 72.2\%. 
However, as an open-source operating system, Android has been attacked by various malware. According to the report released by Qianxin Threat Intelligence Center \cite{report}, a total of 2.3 million samples of malicious programs were intercepted on the Android platform in 2020, with an average of 6,301 new samples of malicious programs every day. 
The AdbMiner Mining Trojan family is active in attacks, capturing tens of thousands of Internet of things devices worldwide, and the number of Internet of things devices captured in China is close to 1000.
Therefore, Android malware becomes so serious that many researchers endeavor to seek effective detection methods.

The advent of machine learning had a significant impact on Android malware detection for the classification stage. Currently, advanced Android malware detection approaches can be categorized into static analysis \cite{Kouliaridis2020} \cite{P.Yan2018} and dynamic analysis \cite{Kouliaridis2020} \cite{Papamartzivanos2014}.
Some researchers utilize state-of-the-art machine learning models like deep learning \cite{IRAM2020}, online learning \cite{Narayanan2017} or ensemble learning \cite{Mantoo2020} to identify multi-class attacks effectively in the Android environment. 

Eliminating redundant or irrelevant features is a significant procedure of machine learning. Babaagba et al. \cite{Babaagba2019} demonstrated the influence of feature engineering in Android malware detection by contrasting the performance of the model with or without feature selection algorithms applied.

As the commonly applied feature selection approach, filter-based Android feature selection models \cite{Salah2020} \cite{Yildiz2019} \cite{Zhang2021} are unable to utilize the feedback from the accuracy of the classifier in Android malware detection, consequently the correlation information between different features obtained from the classifier is ignored. However, the number of possible combinations of these features is so large that it is not infeasible for an exhaustive search in a wrapper-based approach \cite{huda2016}, which always incurs a high computational expense.

In this paper, a wrapper-based feature selection model using DDQN \cite{Mnih2015}, DroidRL, is proposed to automatically select valid Android feature subsets. The main contributions of this paper are summarized as follows:

(1) Reinforcement learning (RL) is leveraged in the wrapper-based feature selection to address the problem of inexhaustible feature subsets of the raw Android features.
Reinforcement learning and exploration-exploitation policy are utilized in DroidRL to explore an optimal feature subset for malware detection.

(2) DroidRL presents an extensible prototype of a feature reduction algorithm for machine learning in other scenarios.
A highly efficient approach is proposed in this paper for researchers to reprocess raw features on their datasets when machine learning models are used. DroidRL takes advantage of the reinforcement learning nature to automatically perform feature selection for dimension reduction, sufficient to replace the burdensome manual feature engineering in the malware detection task.

(3) The feature dimension is notably reduced (1083 to 24) using the DroidRL framework while maintaining high accuracy (95.6\%).
Extensive experiments demonstrate that the DroidRL framework performs better than the traditional feature selection methods, improving detection performance on a variety of classifiers.

The paper is structured as follows:
Section \ref{Related_Work} gives an introduction to related works on feature selection for Android malware and reinforcement learning in cyber security. 
Section \ref{DroidRL} describes the fundamental principles of the DroidRL applied in feature selection in Android malware detection.
Section \ref{Training} introduces the training process of the DroidRL.
Section \ref{Setup} presents the dataset to carry out the comparative experiment, the feature extraction method, and the data preprocessing process. 
Section \ref{Results} discusses the results of the experiment.

\section{RELATED WORK}\label{Related_Work}
\subsection{Feature Selection for Android Malware Detection}

Feature selection is the process of selecting a subset from the original feature set to improve detection efficiency. Determined by whether independent of the accuracy of the classifier, feature selection algorithms can be categorized into filter and wrapper based algorithms. Table \ref{category_table} depicts the traits of each category and the difference between them. 
Yu and Liu \cite{Yu2003} introduced a correlation-based filter (FCBF) feature selection which made improvements in the traditional filter-based approach for reducing the redundancy among relevant features.
Priya et al. \cite{Priya2020} detected Android malware using an improved filter-based technique; the k-nearest neighbor (KNN) based Relief algorithm.
Huda et al. \cite{huda2016} applied the filter's ranking score in the wrapper selection process and combined the properties of wrapper and filters with API call statistics to detect malware based on the nature of infectious actions.
Xu et al. \cite{ICCDetector} employed correlation-based feature selection (CFS)\cite{2000Correlation} to identify and remove the redundant features, reducing 121,621 to 5,000 features.
Yuan et al. \cite{DroidDetector} utilized only features that deep learning essentially exploited to characterize malware and reached 96.76\% detection accuracy.
Allix et al. \cite{BasicBlocks} created basic blocks as features, sequences of instructions in the control-flow graph with only one entry point and one exit point, thus representing the smallest piece of the program that was always executed altogether

\begin{table}[ht]
    \begin{center}
        \caption{\label{category_table}\textbf{Feature subsets selection in Android}}
        \begin{tabular}{
            m{0.08\textwidth}<{\centering}|m{0.2\textwidth}<{\centering}| m{0.3\textwidth}<{\centering}|m{0.35\textwidth}<{\centering}
        }
        \hline
        Method& Advantage& Disadvantage& Algorithm\\ 
        \hline
        Filter&  
        Fast, lower computational cost& 
        Without considering feature relevance&
        Correlation-based Feature Selection (CFS),The Consistency-based Filter \cite{Dash2003},Information Gain \cite{Mcwilliams2014},ReliefF \cite{Spolar2013}\\ 
        \hline
        Wrapper&
        Capture feature relevance, optimize the predictor& 
        High computational cost&
        FFSR \cite{Xiang2010},WrapperSubsetEval \cite{Witten2011}\\ 
        \hline
        \end{tabular}
    \end{center}
\end{table}

The typical filter-based feature selection algorithm is ranking-based. Each feature is assigned a score according to its importance and then the top N features are selected as input for the classification stage after ranking all the features. 
Huang et al. \cite{Huang2008} proposed a parameterless feature ranking approach for feature selection and a modified greedy feature selection algorithm. 
Wang et al. \cite{Wang2014} ranked individual permissions based on the risk of single permission and the group of permissions. 
Mahindru, Arvind, and Sangal \cite{Mahindru2020} applied six different feature ranking approaches to select significant features, including Gain-ratio feature selection, Chi-Square, Information-gain, and logistic regression analysis. 
In another experiment \cite{Mahindru2021} they combined six distinct kinds of feature ranking and four distinct kinds of feature subset selection approaches to select the valid feature subsets. 

Traditional machine learning models can be optimized to select valid feature subsets.
Youn \cite{Youn2002} presented an algorithm based on support vector machines (SVM) for feature selection to decrease the computation time. 
Priya, Varna, and Visalakshi \cite{Priya2020} proposed KNN based relief algorithm for feature selection. The optimized SVM algorithm was applied for malware detection with the result equivalent to the performance of neural network. 

The state-of-the-art machine learning algorithms like the genetic algorithm \cite{Fatima2019} and neural network \cite{Wang2020} are also used in feature subsets selection.
For neural networks, the score derived from the sum of the softmax weights of the input features can be adopted as an evaluation indicator to select valid feature subsets.

From the above discussion, the conclusion can be reached that little research utilizes the feedback from the accuracy of the classifier in Android malware detection. Filter-based feature selection is less computationally expensive compared with wrapper-based feature selection, but the relevance between different features is ignored, which can consequently choose a large number of redundant features while processing high-dimension feature vectors.
To make a improvement in the efficiency of the malware detection classifier, the problem of inexhaustible feature combinations in selected valid subsets in the previous wrapper-based method should be addressed.

\subsection{Reinforcement Learning in Cyber Security}
The prevailing algorithms of reinforcement learning are Q-learning \cite{Melo2001}, Deep Q Network (DQN) \cite{Mnih2015} and Double Deep Q Network (DDQN) \cite{Hasselt2015}. 
DQN is introduced by Mnih et al. to address the problem of the difficulty to use Q-table for high-dimensional, continuous state and action space.
DDQN made a noteworthy improvement on DQN in the training algorithm. The generation method of target Q value is modified in DDQN \cite{Hasselt2015} to deal with the overestimation of the Q value of action in traditional DQN.

Applications of reinforcement learning in software security achieved significant improvements in recent years.
CyberBattleSim \cite{cyberbattlesim} implemented a automated defender agent that detected and mitigated ongoing attacks based on pre-defined probabilities of success, with the simulation environment parameterized by fixed network topology and a set of predefined vulnerabilities.

Especially in virus detection, reinforcement learning has been commonly applied in malware classification \cite{Binxiang2019} or adversarial sample generating \cite{Fang2019} \cite{Rathore2020}.
Fang et al. \cite{Fang2019} trained an AI agent to automatically generate adversarial samples by rewarding it if the modified malware escaped the classifier detection.
Rathore et al. \cite{Rathore2020} generated malware using reinforcement learning for maximizing fooling rate while making minimum modifications to the Android application.
To address the problem of the slow learning rate in the game with the high dimension of Q-learning, Wan et al. \cite{Wan2017} applied deep Q-network technique with a deep convolutional neural network in mobile malware detection, which initiates the quality values based on the malware detection experience.

In this work, by combing the advantages of using existing experience while automatically exploring other optimal subsets in reinforcement learning and the utilization of feature relevance in wrapper-based feature selection, DroidRL tackled the problems of the traditional feature selection algorithm to make the feature selection phase faster and the malware classification more efficient.

\section{DROIDRL FRAMEWORK}\label{DroidRL}

For the DroidRL framework, the primary task is to train a learning agent to sequentially choose valid Android feature subsets by interacting with the environment and utilizing its learned knowledge. This section describes how the DroidRL framework achieves its goal.

\subsection{Overview of DroidRL}\label{Overview_of_DroidRL}
\begin{figure}
    \centering
    \caption{DroidRL for feature selection in Android malware detection}
    \includegraphics[width=0.7\textwidth]{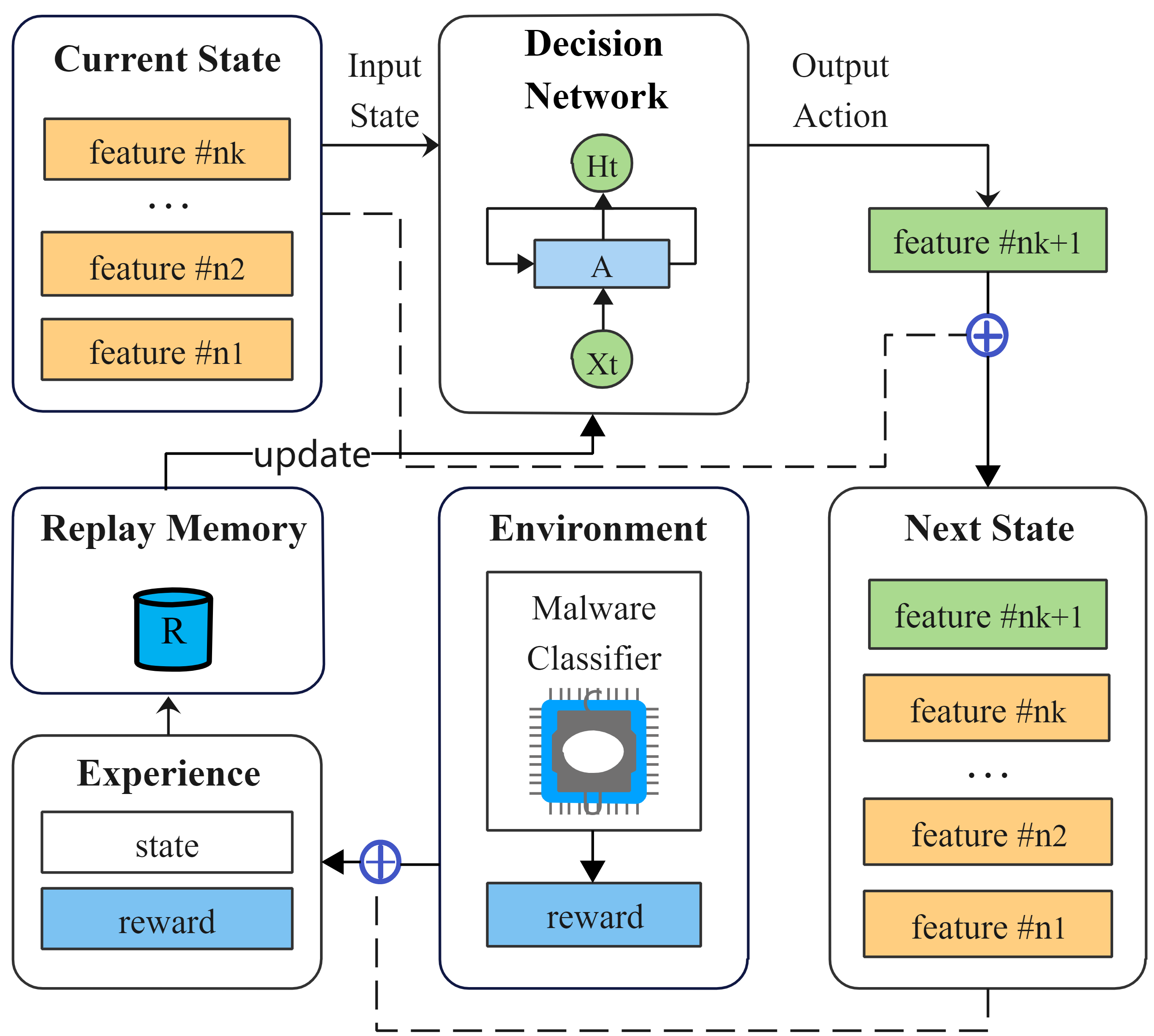}
    \label{RLDroid}
\end{figure}

Figure \ref{RLDroid} shows the schematic diagram of the DroidRL. The core part of the DroidRL framework is built up by the DDQN-based decision network.
In each step, the autonomous agent independently carries out an action decided by the decision network, to select one feature into its observed state from the environment using their prior knowledge.
To evaluate the quality of the feature subset and differentiation of the selected individual features, the reward of the action produced from the malware classifier is determined by the malware classification accuracy using selected features as input.
Furthermore, the state of the agent, the chosen action, and the reward of this epoch are saved in the replay memory for training the decision network. The exploration-exploitation policy is enhanced to address the problem of computational expense due to inexhaustible feature subsets.

\subsection{Key Components in Reinforcement Learning}

(1) \textbf{Environment}: The environment is the place for the agent to explore and get feedback.

In our framework, the environment contains all candidate features and is responsible for putting the agent's current state into the malware classifier after each action is executed. The accuracy of the classification will be returned to the agent as reward.
The total number of given features and the length of the valid feature subsets that need to select are defined in environment. 
When the agent has selected enough features according to the declared length of the valid feature subset, the environment instructs the agent to end this round, return the final reward, and reset itself.

(2) \textbf{Action}: 
 Action is the critical step the agent in reinforcement learning needs to take from action space based on its experience and current state.

In the DroidRL framework, action space contains features in the raw feature sets extracted through decompiled APK files.
The state of the agent describes the currently selected features as the result of a series of actions. 
The primary task of the agent is to find the optimal feature subset that is highly distinguishable between malware and benign Android software.

For each action in the DroidRL framework, one unselected feature is added to the state. $ \varepsilon $-Greedy algorithm is employed to make the agent trade-off between exploration and exploitation. Each action is explored with the probability of $ \varepsilon $, while the action with the largest Q value is exploited with the probability of 1 - $ \varepsilon $. Aimed to enable the agent in the training phase to explore more at the early stage and exploit more using the existing experience at the later stage, some improvements are made to $ \varepsilon $-Greedy algorithm as displayed in Equation \ref{random_probability}, where $episode$  is the current training round, $ E $ is the total training round, and $ P $ is a probability parameter between 0 and 1. More details are given in section \ref{Training}.

\begin{equation}
	\label{random_probability}
    \varepsilon = 1-\frac{episode}{E}\times p
\end{equation}

(3) \textbf{Reward}: The reward is feedback as a result of interaction between the agent and the environment through taking action.

In this paper, the reward is determined by the accuracy of the Android malware classifier, with the selected features in agent's current state as input. The agent enters a state $ s $ after executing feature selection action $ a $. Then the reinforcement learning environment returns the corresponding reward from the malware classifier to evaluate the action value function Q.

With the goal to obtain the highest Q value by action, the agent can consequently find the valid feature combination devoted to the highest accuracy.
In Equation \ref{Q_1}, $s$ and $a$ respectively represent the current state and the action taken at the current step. $r$ is the obtained reward, $s'$ represents the next state reached by the agent, and $a'$ refers to the action that can obtain the highest Q value in the next state. 
As $a'$ can also be calculated by Q function, the original formula is equivalent to Equation \ref{Q_2}, where $\theta_{1}$ and $\theta_{2}$ represent the parameters of two networks in DDQN respectively. 

\begin{equation}
	\label{Q_1}
    Q_{(s,a)} = r + \gamma Q(s', a')
\end{equation}

\begin{equation}
	\label{Q_2}
    Q_{(s,a,\theta_{1})} = r + \gamma Q(s', \mathop{\arg\max}_{a}Q(s',a, \theta_{2}), \theta_{1})
\end{equation}

\subsection{Decision Network}
In the DroidRL, the decision network is the brain of the agent. When exploitation is performed, the agent puts the current state represented by a vector into DroidRL's decision network, then the decision network returns the guidance of the next action to the agent. 

It is worth mentioning that the length of the agent's state is continuing to grow. It causes the input of the decision network to have an unfixed length. DroidRL's decision network needs to be specially designed since the input to a neural network is normally in constant shape.
RNN is a kind of neural network widely used in natural language processing. Because of the unfixed length of natural languages, RNN-like networks are designed to accept input of indeterminate length. For this reason, DroidRL's decision network adopts RNN and its variants.

The DroidRL also applies some tricks that can do help to the training effect and the ability of the feature selection.  

\textbf{(1)Word Embedding:}
The input of our decision network is a sequence of selected features.
Instead of being presented as the one-hot vector, word embedding is applied to process the input.
If the one-hot vector is used to represent the sequence of the features, the entire input matrix will be large and sparse. It will lead to a huge amount of computation and storage. Additionally, there is no semantic information when features are represented by one-hot vectors, which is not conducive for the decision network to find the correlation between features.
Applying word embedding into DroidRL's decision network can improve the framework in the following two aspects:
\begin{enumerate}
\item [1.] Compressing a one-hot vector into a denser one. Word embedding greatly reduces the input dimension and improves the training speed of the model. 
\item [2.] Compressed feature vectors are more semantic. The DroidRL's main task is to select an optimal subset of features. After word embedding is added to the decision network, the DroidRL can cluster features in high dimension space according to their semantics. Then it can better find the features that can be combined with the current selected features.  
\end{enumerate}

\textbf{(2)Features Ordering:}
There is a special consideration for applying natural language processing methods to DroidRL feature selection. Natural language is sequential in nature which means replacing two words in a sentence can make the sentence confusing and meaningless.
However, in feature selection, replacing any two selected features in the feature sequence should not have any influence on the decision network making its decision. The input state [1,2,3] and the input state [1,3,2] are identical in meaning since they contain the same features and should produce the same output in the decision network. Exchanges positions of any two features in the input state should not influence the result. 
This character of feature selection differs from that of natural language. If this particular property is not addressed, it may have a negative impact on RNN-like decision network learning.  

A trick is applied to deal with this problem. Before the features are fed into the decision network, they are sorted by index. In this way, the same feature set can be guaranteed to produce only one input regardless of the order of feature selection.

\begin{figure}
    \centering
    \caption{\textbf{DroidRL's Decision Network}. The decision network takes in the agent's current state and predicts a new feature as action.}
    \includegraphics[width=0.85\textwidth]{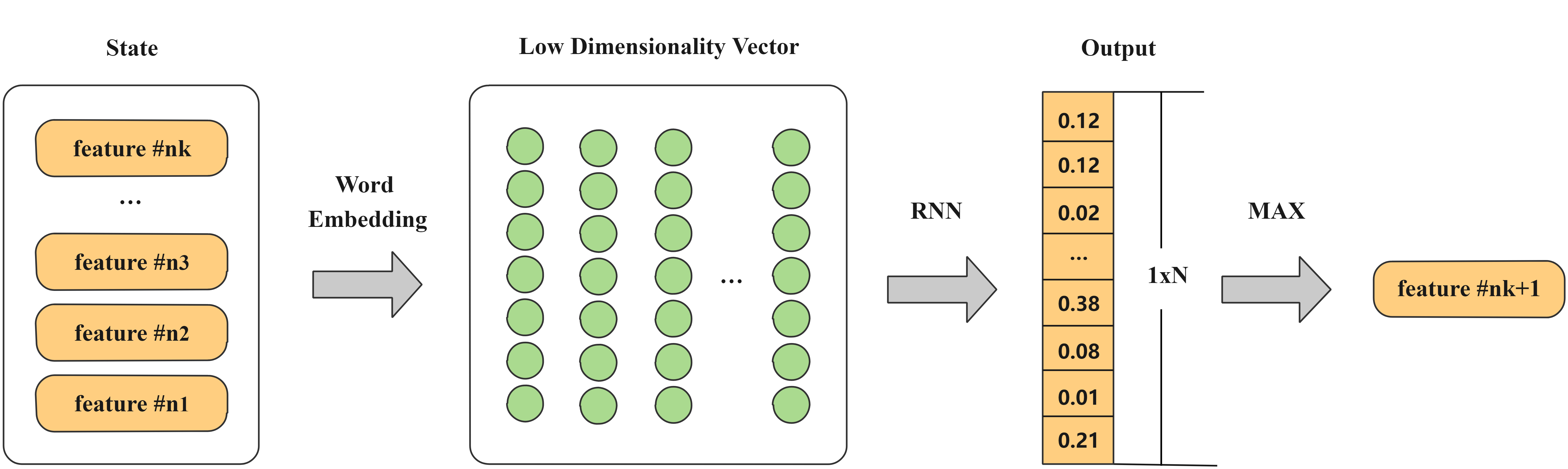}
    \label{fig:network}
\end{figure}

After applying the above tricks, the ultimate decision network structure is shown in Figure \ref{fig:network}. The agent puts its state, a sequence that represents the selected features, into the decision network. The first layer of the decision network is the embedding layer. Features represented by one-hot vectors go through the embedding layer and become more dense vectors. These vectors then are fed into the RNN-like network and finally enter a fully connected layer and a softmax layer. 

\section{TRAINING PHASE}\label{Training}

The DroidRL training process is elaborated in this section. The training algorithm and evaluating algorithm are illustrated in Algorithm \ref{algo1}.

\begin{algorithm}
  \caption{training and evaluation process}\label{algo1}
  \SetAlgoLined
  \KwData{length of feature dictionary $N$, number of features to be selected $F$, total training episode $E$, replay memory warm-up episodes $M$, initial replay memory $rmp$, initial network weights $\theta_{1}$,$\theta_{2}$, initial probability of exploration $\varepsilon$ }
  \KwResult{Optimal feature subset for malware detection}
  initialization\;
   \For{$m \leftarrow 1$ \KwTo $M$}{
        sample some random states and feed into $rmp$\;
   }
   \For{$episode \leftarrow 1$ \KwTo $E$}{

        $state$ $\leftarrow$ {an\ empty\ array}
        
          \For{$f \leftarrow 1$ \KwTo $F$}{
          $previous$\,$state$ $\leftarrow$  $state$;
          
        with probability $\varepsilon$ random choose a feature and put it into the $state$\;
        
        or
        
        input the state to Decision network and get an N dimension vector, find the index of the max value in the vector and put it into the $state$\;
        
        calculate $reward$ of the $state$\;
        put $reward$, $previous$\,$state$, $state$ into $rpm$;
        }
        
    \If{$episode$ mod $Network\_Learn\_Frequency$ == 0}{
      sampling some samples from $rpm$ and update $\theta_{1}$\;
      }

    \If{$episode$ mod Sync\_Frequency == 0}{
      sampling some samples from $rpm$ and learning\;
      Synchronize $\theta_{1}$,$\theta_{2}$\;
      }
       update $\varepsilon$  by equation \eqref{random_probability}\;
        
   }
   
  start evaluating\;
  
    $optimal$\,$state$ $\leftarrow$  an\,empty\,array\;
     \For{$f \leftarrow 1$ \KwTo $F$}{
          input the current $optimal$\,$state$ to Decision network and get an N dimension vector, find the index of the max value in the vector and put it into the $optimal$\,$state$\;
    }
    recalculate $final\_reward$ of the $optimal$\,$state$\;
    \textbf{return} $optimal$\,$state$, $final\_reward$\;
\end{algorithm}

To overcome the problems of correlated data and non-stationary distribution of training data, replay memory is adopted in DroidRL. Before training begins, DroidRL obtains some initial samples by running warm-up episodes and feeds them into the replay memory. At the beginning of each training episode, the state of the agent is cleared and then the agent begins to select features. Each episode ends after a sufficient number of features have been selected. During the training episodes, a strategy that traded off between exploration and exploitation is adopted. As depicted in equation \eqref{random_probability}, the agent has a very high probability of exploration at the beginning but more possibility of exploitation as the increase of episode.

In the case of exploration, the agent randomly selects a feature (that is not in its state) in an action. Exploration allows the agent to try more possible feature combinations and choice space. 

When the exploitation policy is executed, the agent uses previous experience to select the optimal feature. Instead of randomly selecting a feature, the agent puts its current state into the decision network and gets a vector of the same length as the feature dictionary that denotes the confidence of each feature. The agent takes the feature with the highest confidence as action. If the highest confidence feature has already been selected, the agent takes the second highest one instead, and so on. As mentioned before, after each time a new feature is added to the state, the state is reordered to ensure consistency of feature set representation. After each action is taken, the features in the state are used to classify malware and benign. The accuracy of the classification as the reward combines with the previous state and the current state after taking the action are put into the replay memory. The agent continues exploit and explore until enough features are selected. 

After training all training episodes, the agent will finally run an evaluation episode. In this episode, each step of agent will utilize what it has learned during the training phase. The output of this episode is the final optimal feature subset.

\section{EXPERIMENT SETUP}\label{Setup}
\begin{figure}
    \centering
    \caption{Overview of DroidRL}
    \includegraphics[width=0.63\textwidth]{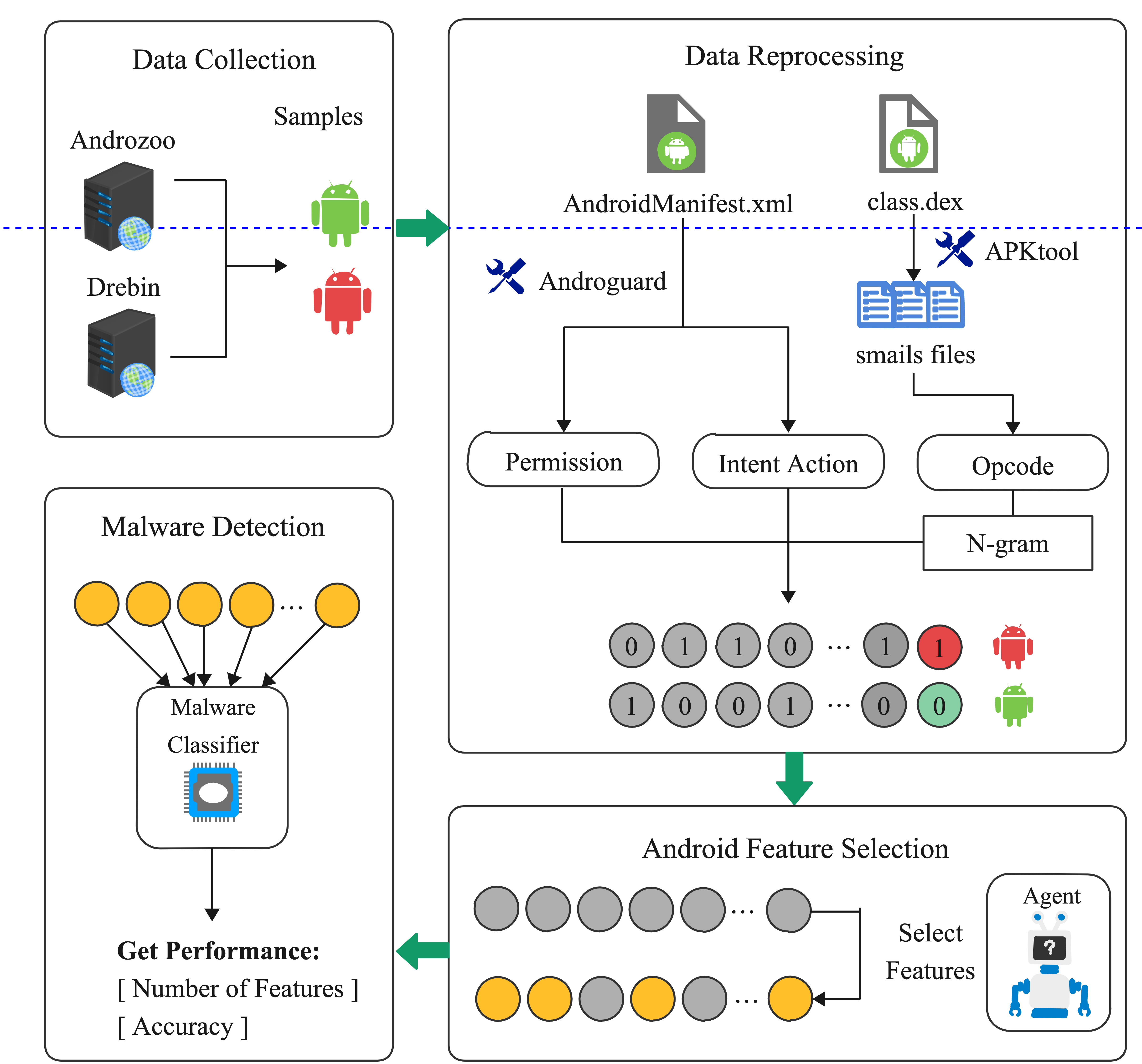}
    \label{process}
\end{figure}
This section provides information on the hardware environment for training DroidRL, our dataset, and hyperparameters setting.

\subsection{Training Environment}
We performed all the experiments on the server with single Tesla V100 GPU and a CPU with two cores. The GPU was used to accelerate the training of decision network in DroidRL, while the training and prediction of the classifiers in DroidRL used CPU only. After the training process, DroidRL's classifier can be extracted separately for testing or deployed on any hardware that capable of running machine learning algorithms.

\subsection{Dataset}
DroidRL's dataset contains 5000 benign samples from AndroZoo and 5560 malware from Drebin to train and test the model. Both data sources are universally utilized in recent years' research focusing on Android malware detection, which renders it easier to carry out comparison experiments with other feature selection methods.
AndroZoo \cite{Allix2016} updates the collection of 16,000k different APKs from several sources including Google Play, with each application analyzed by different AntiVirus products to label the Malware. 
Malware samples in this research are mainly selected from Drebin \cite{Daniel2014}, a commonly used dataset that contains 5,560 applications from 179 different malware families. 

Static analysis is applied in this work, extracting the permissions, intent actions, and opcode as original features from android samples decompiled by APKtool and Androguard for further reinforcement learning based feature selection. 

In total 457 permissions and 126 intent actions that are typically considered to be highly relevant to the malicious behavior of Android applications, are chosen in this paper to construct original feature set.
Permissions indicate what sensitive user data (e.g., contacts and SMS) need to be accessed by an application, essential in Android malware detection. Intent actions are abstract objects containing information on the operation to be performed for an app component.

After disassembling the class.dex to generate the smalis files, Dalvik bytecode (example: invoke-direct) is gained through scanning the method field in smalis files with regular expression. Opcodes are obtained by mapping the Dalvik bytecode to a series of letters as described in Table \ref{opcode_table}.

\begin{table}[H]
    \begin{center}
        \caption{\label{opcode_table}\textbf{Dalvik Instruction Transformation Table}}
        \begin{tabular}{m{0.1\textwidth}<{\centering} m{0.8\textwidth}<{\centering}}
        \hline
        \textbf{Letter}& \textbf{Dalvik Instruction} \\ 
        \hline
        M & 
        move, 
        move/from16, 
        move/16, 
        move-wide, 
        move-wide/from16, 
        move-result, 
        move-wide/16, 
        move-object, 
        move-object/from16, 
        move-object/16 ...\\ 
        \hline
        R &
        return-void, 
        return, 
        return-wide, 
        return-object\\
        \hline
        G &
        goto, 
        goto/16, 
        goto/32\\
        \hline
        I &
        if-eq, if-ne, if-lt, if-ge, if-gt, if-le, if-eqz, if-nez, if-ltz, if-gez, if-gtz, if-lez\\
        \hline
        T &
        aget, aget-wide,
        aget-object,
        aget-boolean,
        aget-byte,
        aget-char,
        aget-short,
        iget,
        iget-wide,
        iget-object,
        iget-boolean,
        iget-byte,
        iget-char ...\\
        \hline
        P &
        aput,
        aput-wide,
        aput-object,
        aput-boolean,
        aput-byte,
        aput-char,
        aput-short,
        iput,
        iput-wide,
        iput-object,
        iput-boolean,
        iput-byte,
        iput-char...\\
        \hline
        V &
        invoke-virtual,
        invoke-super,
        invoke-direct,
        invoke-static,
        invoke-interface,
        invoke-virtual/range,
        invoke-super/range,
        invoke-direct/range...\\
        \hline
        \end{tabular}
    \end{center}
\end{table}

Opcode features are segmented by N-gram to obtain the transformation sequence. 
The dimensionality reduction approach is employed to address the problem of the high dimensionality of the feature vector due to the increase of the number of N-grams with the value of N.
Firstly, the N-gram set of malicious samples is extracted by the N-gram extraction process proposed in \cite{ZHANG2019}. The top k high-frequency N-grams are selected, presented as $Set=\left \{x_1,x_2,x_3,\cdots,x_k \right \}$. Subsequently, a k-dimensional binary feature vector $feature=[m_1,m_2,m_3,\cdots,m_k]$ is constructed for the sample based on this feature set, where $m_i$ is "1" indicates that the N-gram set of the sample contains the element $x_i$ in the feature set.

From the above discussion, the whole process to detect Android malware using DroidRL for feature selection is depicted in Figure \ref{process}. 
Android application samples are collected from Drebin and Androzoo datasets, decompiled by Android reverse engineering tools to extract the permissions, intent actions, and N-grams as the original features. Then DroidRL feature selection is applied to select the valid feature subsets from the original features. The valid feature subset is saved and employed to validate the performance of the malware classifier, using the number of features and accuracy as evaluating index.
10-fold cross-validation is used in the experiment to evaluate the models and avoid overfitting. The features selected by DroidRL will be used to train a final classifier for malware detection.

\subsection{Hyperparameters setting}
The detailed description and setting of DroidRL's parameters are shown in Table \ref{hyperparameters}. The classifiers in DroidRL are built with scikit-learn and all hyperparameters used are set by default. 

\section{EXPERIMENT RESULTS}\label{Results}
The following research questions have been brought out to help follow the process of experiment conduction:

\textbf{RQ1}
Is the result produced by the DroidRL framework stable selecting only a dozen features from a high-dimensional exploration space (e.g.1083, the dimensionality of the original feature vector)?

\textbf{RQ2}
Does the decision network learn the key information conducive to the next optimal feature selection in the process of training?

\textbf{RQ3}
What is the performance of different classifiers using the optimal feature subset selected by the DroidRL framework as input?

\textbf{RQ4}
What is the impact on the training time of malware classifiers of using the feature subset selected by DroidRL rather than the original features?

\textbf{RQ5}
How is the performance of the DroidRL framework compared with other advanced methods in related work?

\subsection{Stability Evaluation of Feature Selection Results}
The key to adopting reinforcement learning to explore the best subset of features for Android malware detection is to find the best combination of features. However, with the exploration-exploitation strategy, DroidRL could take different actions even in the same state. There are inevitable differences in the optimal feature subset selected by the DroidRL.
Moreover, the space that can be explored by reinforcement learning is gigantic. Only about 24 of the total set of 1083 features in the experiment are selected as input for malware detection. 
The different results may also cause the instability of the experiment results.

\begin{figure}
    \centering
    \caption{Stability verification of different feature selection results by DroidRL}
    \includegraphics[width=0.63\textwidth]{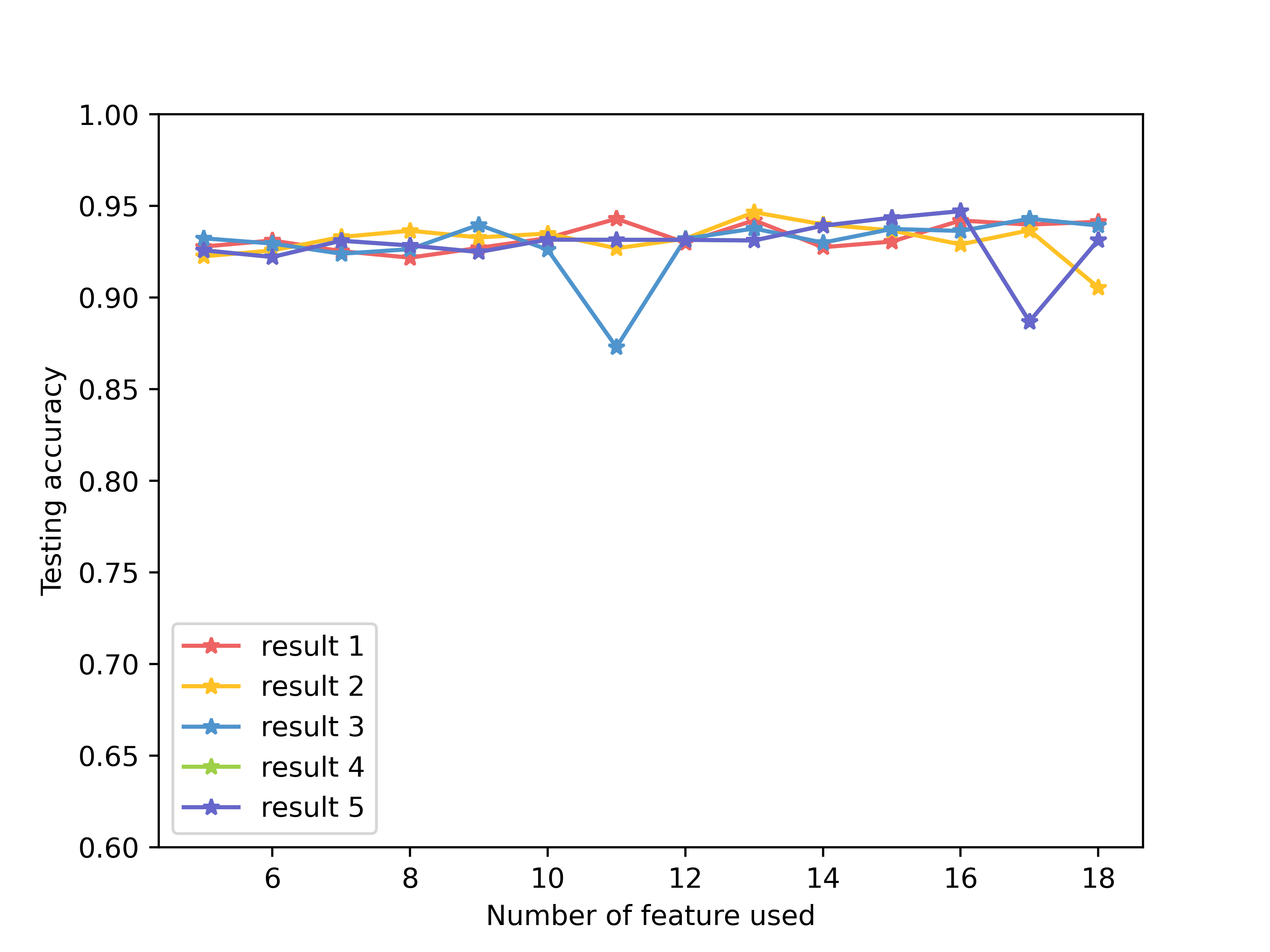}
    \label{fig:eval_obs}
\end{figure}

In an attempt to verify whether the randomness brought by the exploration-exploitation strategy will affect the final detection accuracy, and to further prove the stability of the feature selection result of DroidRL, this experiment uses decision tree (DT) as the classifier and tested the accuracy using five different feature subsets results obtained from DroidRL as input. The result is illustrated in Figure \ref{fig:eval_obs}

Although the final results of the features selected by reinforcement learning in the five experiments are not strictly identical, it can be seen from Figure \ref{fig:eval_obs} that the detection accuracy always lies in the 92\%-95\% range with few points appearing deviation due to the randomness brought by the exploration-exploitation strategy. Though the action taken by the agent each time can not be exactly the same when exploring the inexhaustible combination of feature subsets, the detection accuracy of using the selected features for malware classification generally remains stable.

\subsection{Evaluation of the Learning Procedure of DroidRL}

\begin{figure}[htbp]
    \centering
    \subfigure[Test the DroidRL model after each training episode]{
        \begin{minipage}[t]{0.7\linewidth}
        \centering
        \includegraphics[width=0.7\linewidth]{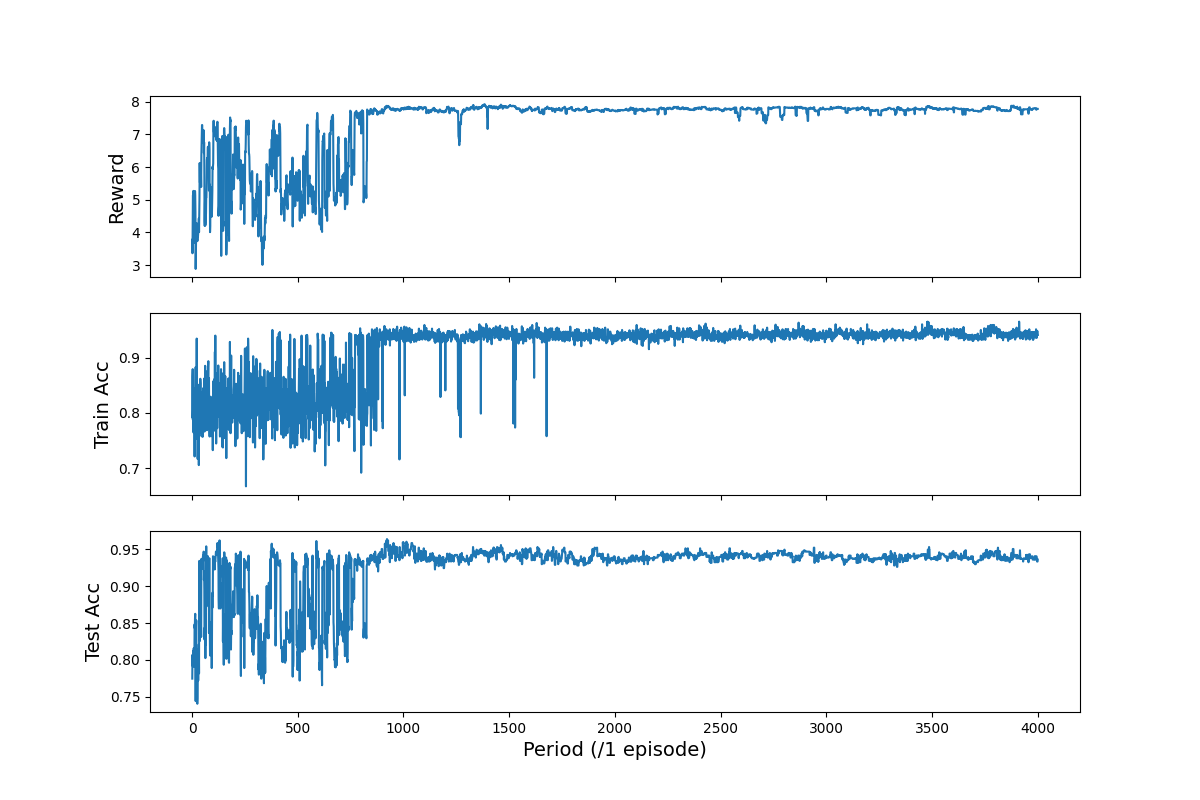}
        \end{minipage}
    }
    \subfigure[Test the DroidRL model after every 50 training episode]{
        \begin{minipage}[t]{0.7\linewidth}
        \centering
        \includegraphics[width=0.7\linewidth]{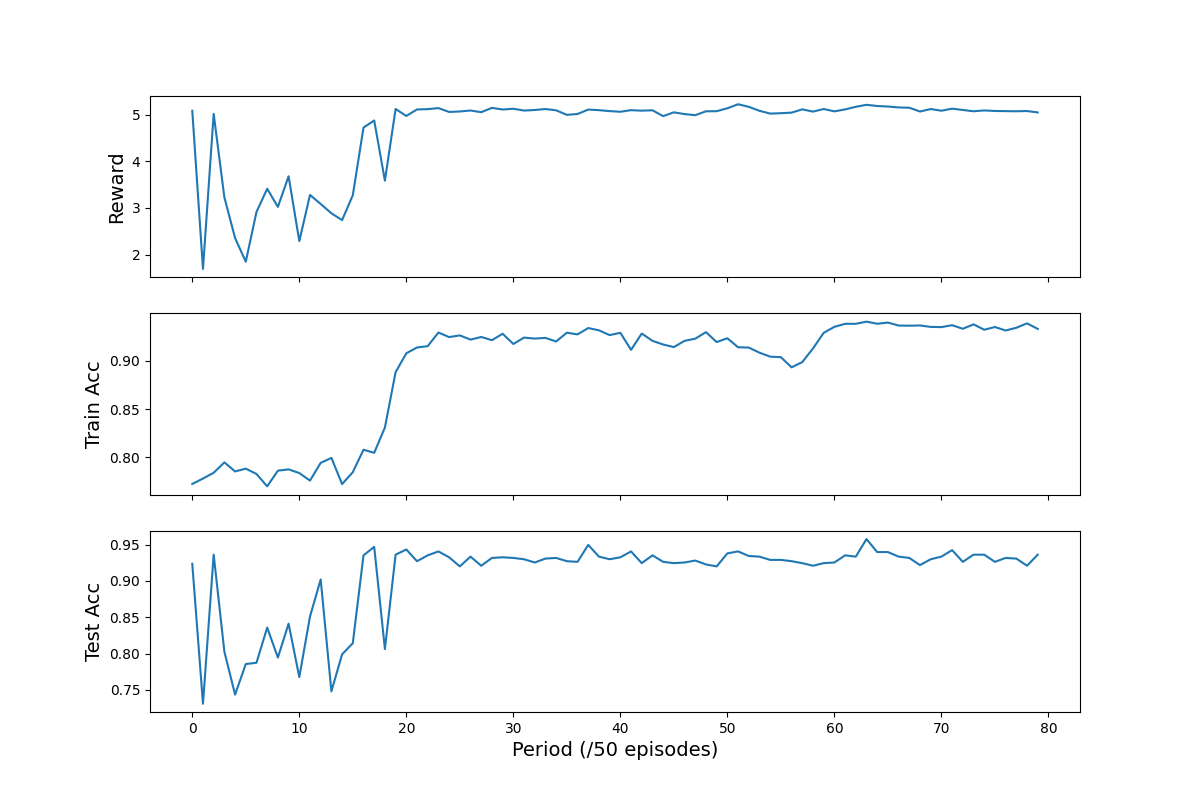}
        \end{minipage}
    }
    \caption{The evaluation of the performance of the DroidRL framework during training}
    \label{eval_network}
\end{figure}

To illustrate the learning procedure of the decision network in the DroidRL framework, the reward, the training classification accuracy (Train Acc in Figure \ref{eval_network}), and the testing classification accuracy (Test Acc in Figure \ref{eval_network}) in each training episode were tracked. 
The accuracy in one episode (e.a., training evaluation episode or testing evaluation episode) is gained from the classifier using the selected features as input.
In the training episode, the agent select features using exploration-exploitation strategy and the decision network is in the training mode; Train Acc is the malware classification accuracy by running one training evaluation episode on the training dataset.
In the testing evaluation episode, the feature selection process of the agent is only guided by the decision network without random exploration; Test Acc is the average of the accuracy in the five testing evaluation episodes.

After each training episode, the DroidRL framework was tested by running five testing evaluation episodes in Figure \ref{eval_network} (a).
In addition, taking 50 training episodes as a period, the DroidRL framework was tested after every period in Figure \ref{eval_network} (b) with the same approach to calculate the testing classification accuracy. Moreover, the training accuracy of one period was recorded as the average malware classification accuracy of the 50 training episodes.

In this experiment, the Long Short-Term Memory (LSTM) \cite{2012Long} severs as DroidRL's decision network since it has better contextual memory than RNN. The Decision Tree (DT) is applied as the classifier 
Similar tendencies are observed both in Figure \ref{eval_network} (a) and Figure \ref{eval_network} (b), as both results display the increasing reward and the malware classification accuracy.
However, significant differences can be witnessed in the two figures, the training classification accuracy fluctuate greatly in Figure \ref{eval_network} (a), but remains comparatively low before 20 periods (e.g., 1000 episodes) and then rises to a higher stable state after in Figure \ref{eval_network} (b).
Also, the testing classification accuracy is always slightly higher than the training classification accuracy.
The above experimental results can be explained as follows:

(1) Both figures indicate the increasing reward and malware classification accuracy, showing learning performance of the DroidRL framework.

(2) In the training episodes of the DroidRL, $\epsilon$-Greedy algorithm was used to balance between exploitation and exploration, but it was inevitable to select some redundant or irrelevant features in the exploration process.
As shown in \ref{eval_network} (a), the fluctuated training accuracy is always higher than the testing accuracy. After about 20 periods, The testing accuracy becomes stable through utilizing existing experience while the training accuracy still has some sudden drops caused by $\epsilon$-Greedy algorithm.

(3) After using the average accuracy of 50 episodes, the training classification accuracy in Figure \ref{eval_network} (b) is more stable in the first 20 periods compared with Figure \ref{eval_network} (a), which clearly shows its changing trend.
Due to the gradually decreasing exploration probability of $\epsilon$-Greedy algorithm in the later stage, the agent tended to use the existing experience from the decision network for feature selection. It led to higher and more stable training classification accuracy, as well as an accuracy curve more similar to the testing phase in the later periods, as shown in Figure \ref{eval_network} (b).

\subsection{Comparison with Different Decision Networks}

\begin{figure}[htbp]
    \centering
    \caption{The evaluation of different decision networks}
    \includegraphics[width=0.63\textwidth]{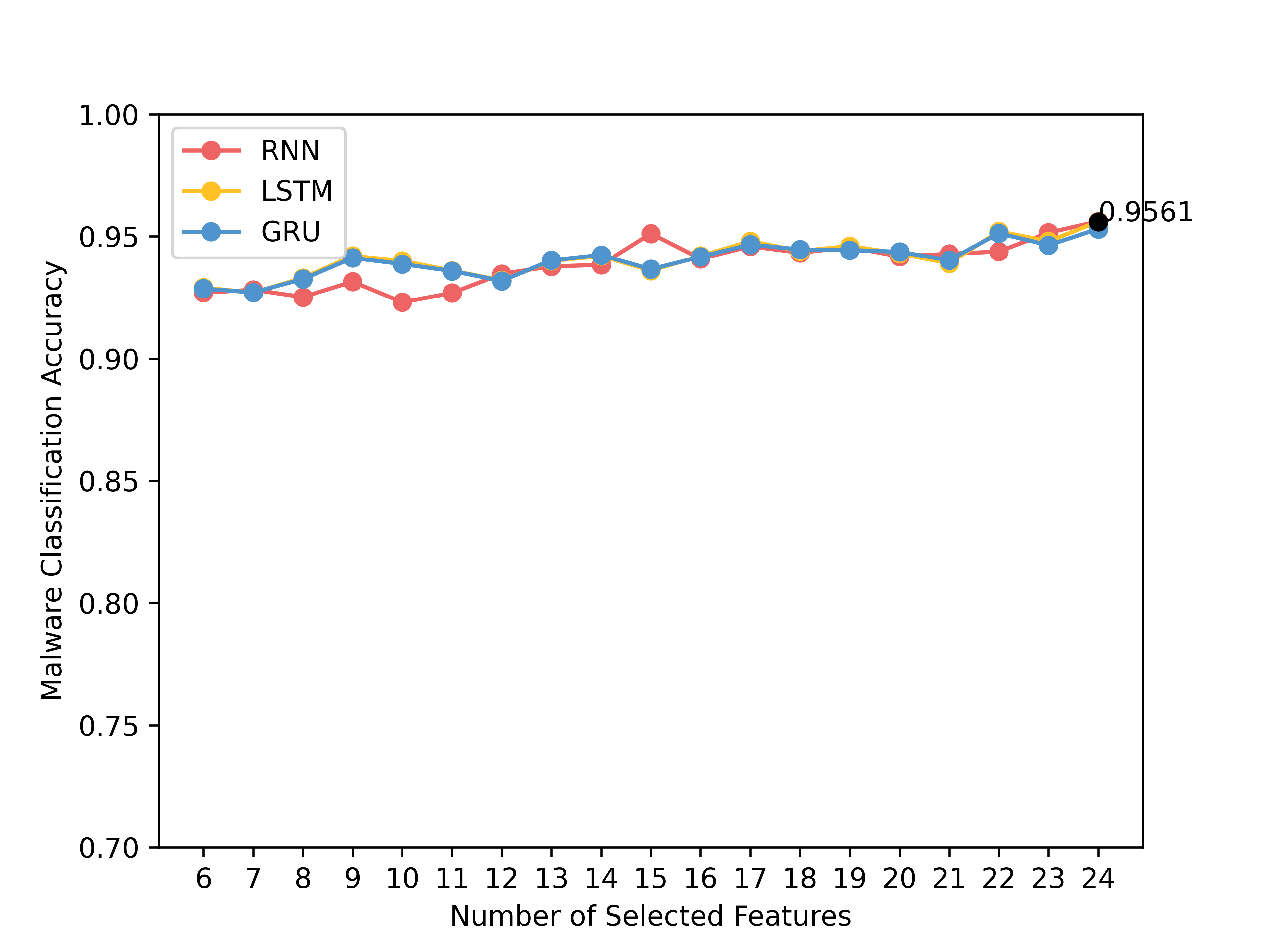}
    \label{eval_obs}
\end{figure}

It is common practice to clarify the decision network selection criteria for a reinforcement learning based algorithm. Therefore, experiments on comparison with different decision networks are conducted and the results are described detailed in this section.

Effective for processing data with sequence characteristics, RNN and its variants have the capability to exploit the temporal and semantic information in the input data and are universally applied to predict the following content according to the context in natural language processing. Therefore, the DroidRL framework adopted an RNN-like network as the decision network to predict the feature to be selected in the next step according to the Android features selected in the previous steps. 

With an attempt to explore which was the most suitable recurrent neural network as the decision network of DroidRL, this experiment applied RNN, Long Short-Term Memory (LSTM) \cite{2012Long} or Gated Recurrent Unit (GRU) \cite{cho2014learning} as the decision network for training respectively. 

In the experiment, 10-fold cross-validation is employed on the shuffled dataset. The average accuracy on the testing set was taken as the classification accuracy. The results are presented in Figure \ref{eval_obs}. 

As the experimental results figures sketch, with more selected Android features in the final valid feature subset, steadily increased higher accuracy is witnessed in RNN, GRU, and SLTM.
The accuracy fluctuation was the most stable when using GRU as the decision network.
The model obtained the best performance (eg. 95.6\% accuracy) with 24 features selected in LSTM as input for malware classification, and the computational overhead was reduced by 97.78\%.

Based on the above experimental results, the following explanations are concluded according to the principle of the DroidRL framework.
Reinforcement learning has a significant effect on selecting the optimal Android feature subset. Adopting the traditional LSTM as the decision network, DroidRL reduced the computational overhead by 97.78\% and retains the accuracy of 95.6\%.

\subsection{Comparison with Different Classifiers}

\begin{figure}[htbp]
    \centering
    \subfigure[DT as classifier]{
        \begin{minipage}[t]{0.7\linewidth}
        \centering
        \includegraphics[width=0.7\linewidth]{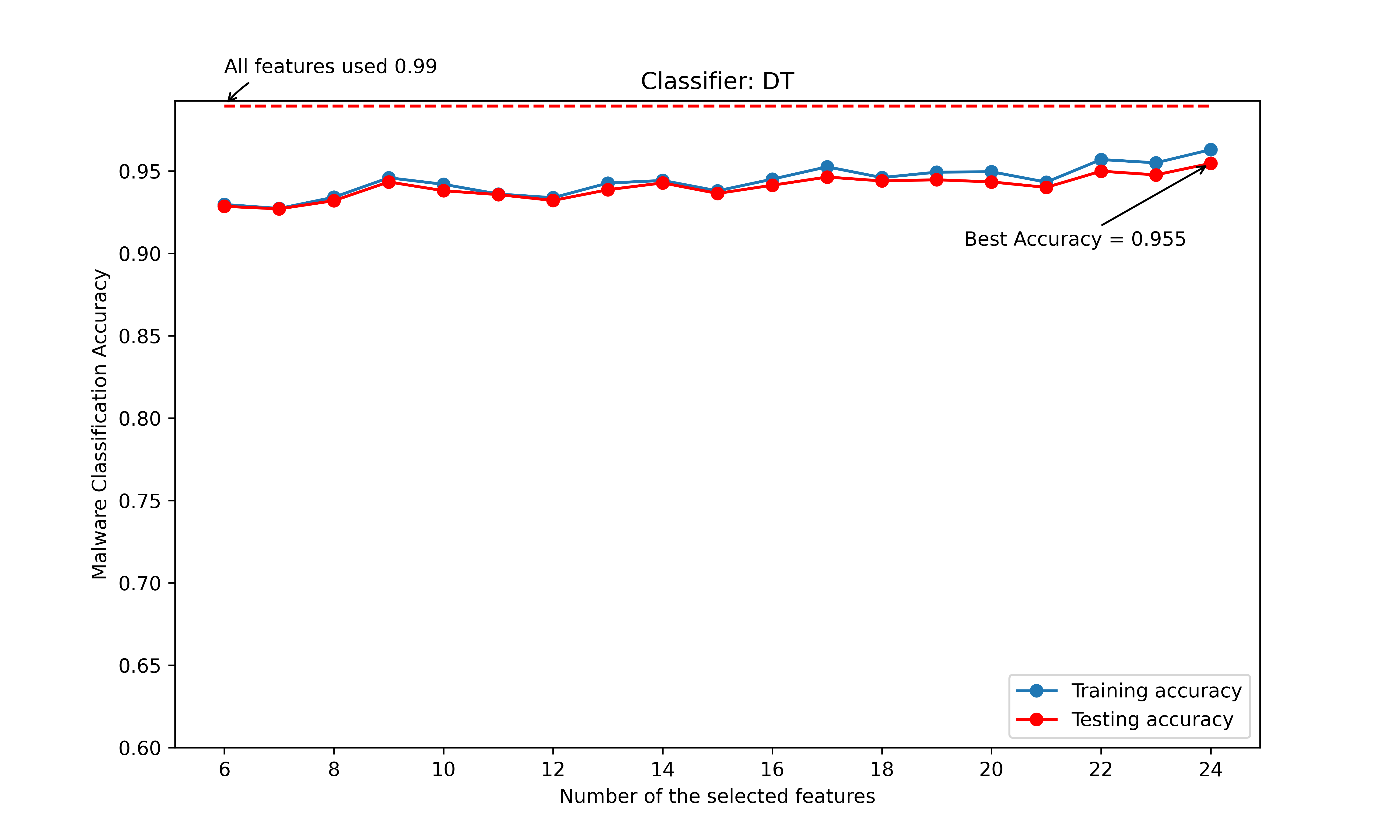}
        \end{minipage}
    }
    \subfigure[RF as classifier]{
        \begin{minipage}[t]{0.7\linewidth}
        \centering
        \includegraphics[width=0.7\linewidth]{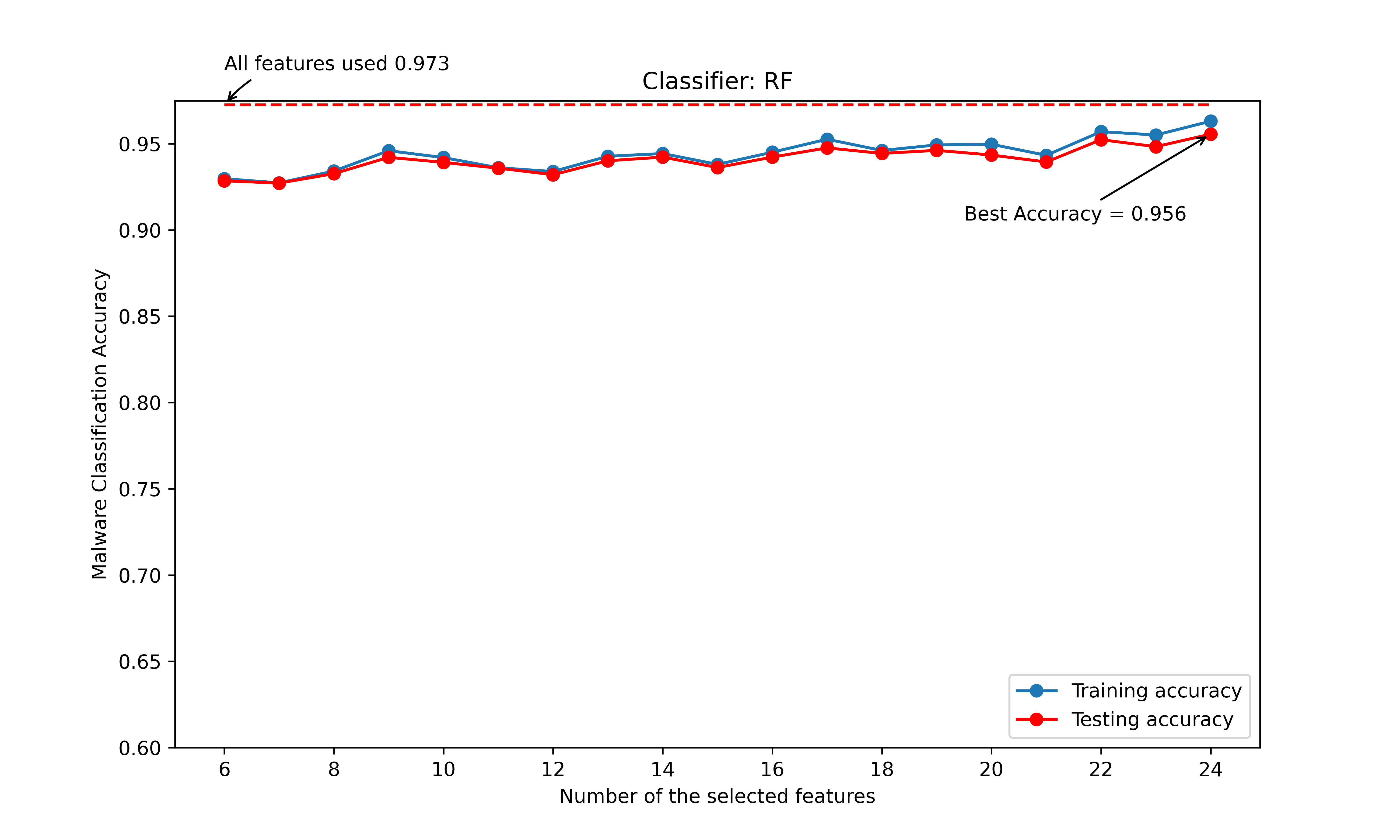}
        \end{minipage}
    }
    \subfigure[SVM as classifier]{
        \begin{minipage}[t]{0.7\linewidth}
        \centering
        \includegraphics[width=0.7\linewidth]{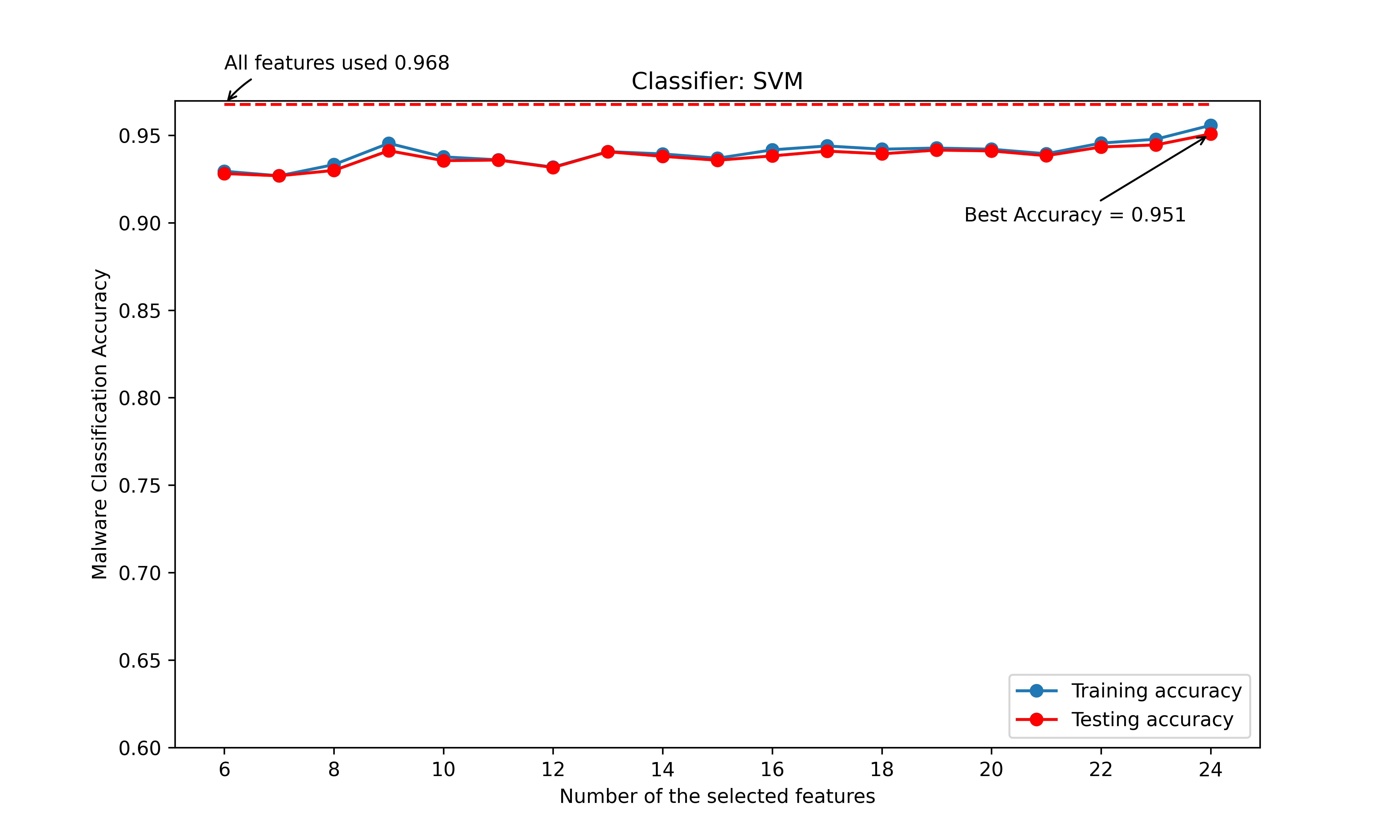}
        \end{minipage}
    }
    \subfigure[KNN as classifier]{
        \begin{minipage}[t]{0.7\linewidth}
        \centering
        \includegraphics[width=0.7\linewidth]{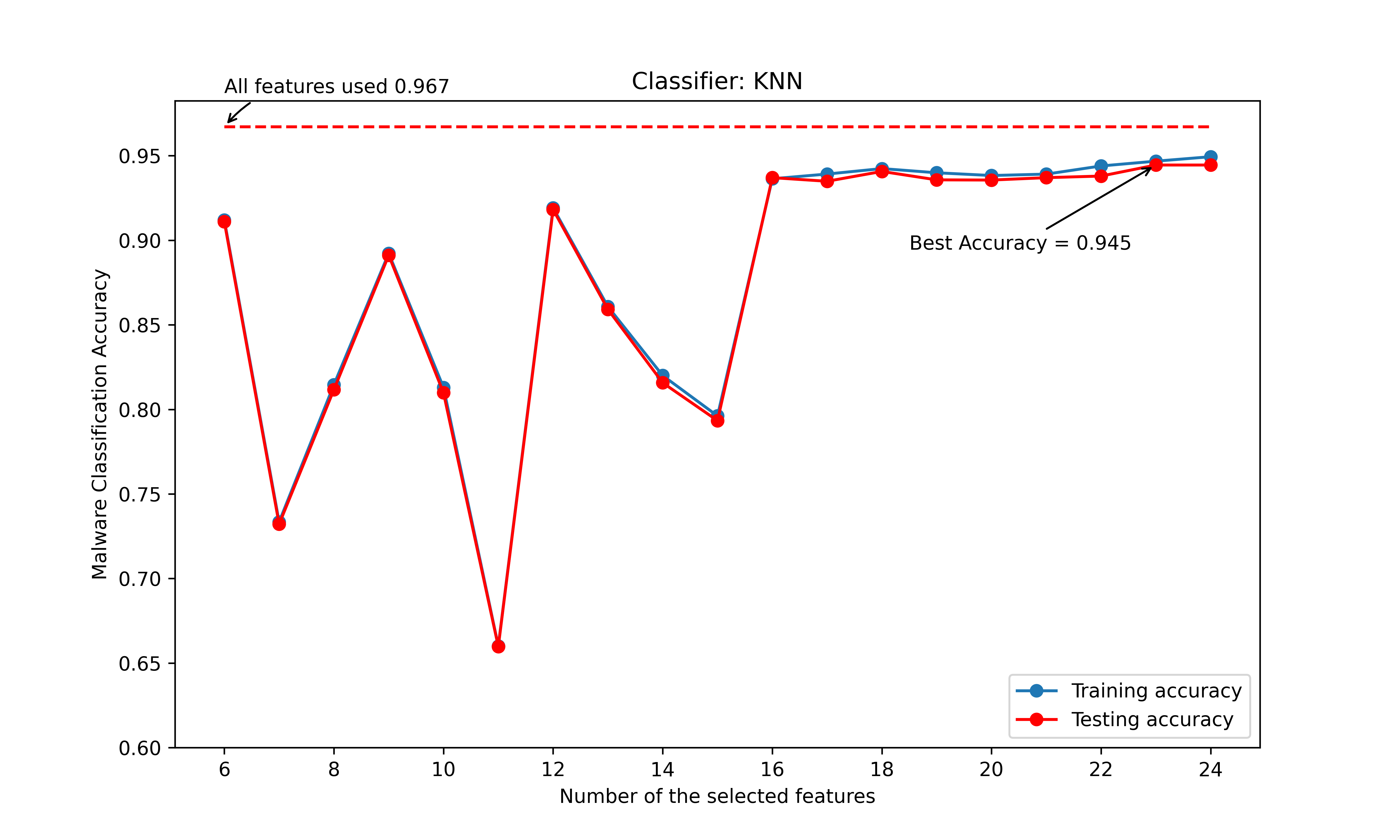}
        \end{minipage}
    }
    \caption{The performance on different classifiers}
    \label{classifiers}
\end{figure}

To demonstrate the performance of our model, we conducted a series of comparative experiments to find the combination of different quantities of features and different classifiers. As shown in Figure \ref{classifiers}, the vertical axis represents the max number of features selected by reinforcement learning-based algorithm for malware detection. More precisely, to make the learning procedure efficient and meaningful, the number of features selected by the model is limited from 6 to 24. 

It can be seen in Figure \ref{classifiers}, when the number of selected features is relatively small, the accuracy is about 90\%. As the number of selected features increases, the accuracy gradually improves, raising to 95\%. 

Figure \ref{classifiers} shows that reinforcement learning can be applied to Android malware detection to select valid feature sets for malware classification.
The accuracy is stable on Random Forest (RF), Decision Tree (DT), and Support Vector Machine (SVM). It can be observed in the Table \ref{performence1} and Table \ref{performence2} that DT, RF, and SVM achieve higher accuracy with fewer features and have a stable performance.
K-Nearest Neighbors (KNN) models perform not so well, as seen from the fluctuation of the accuracy in figures. After the verification experiment, it is found that different hyperparameters k are suitable for classifying different feature numbers, so the accuracy fluctuates greatly with the number of selected features.

\subsection{Comparison of the Classifiers' Training Time}
The computational complexity of machine learning algorithms grows with the number of samples and features.
While the increase in sample size can bring more robustness to the classifier, the increasing number of features could bring redundancy. 
A large number of features also increase the computational complexity and resources required for training.
To evaluate the ability of DroidRL to retain training efficiency, this experiment uses feature subsets with different lengths selected by DroidRL to train various malware classifiers.
To measure the improvement in training efficiency, we calculated the ratio of the time consumed to train the classifier using a subset of features to the time spent using the full feature set.

As shown in Table \ref{time_ratio}, using the subsets to train the models significantly improves the training efficiency. The reason for the relatively large ratio on Random Forest mainly lies in that there are many subtrees in the model. There is a lower bound on the time to train subtrees as the number of features decreases.

Another noteworthy phenomenon is that the training time ratio does not strictly improve with the number of features, and sometimes even drops. This indicates that DroidRL filters out features that are useful for classification, and the added features allow the classifiers to easier find decision boundaries, which in turn speeds up training.

\begin{table}[H]
\begin{center}
    \caption{\label{time_ratio}\textbf{Training time ratio}}
    \begin{threeparttable}
    \begin{tabular}{|l|c|c|c|c|c|c|c|c|c|c|}
    \hline
    \diagbox{\textbf{Classifier}}{\textbf{Ratio(\%) \tnote{1}}}{\textbf{Features Used}}&
    16& 17& 18& 19& 20& 21& 22& 23& 24\\
    \hline
    \textbf{Decision Tree}&
    1.23& 1.64& 1.36& 1.78& 1.63& 1.62& 1.91& 2.06& 2.78\\
    \hline
    \textbf{Random Forest}&
     15.07& 16.64& 15.66& 17.35& 16.64& 16.36& 16.34& 16.85& 17.81\\
    \hline
    \textbf{Support Vector Machine}&
     3.53& 4.22& 4.13& 4.50& 4.51& 4.49& 4.79& 4.74& 4.88\\
     \hline
    \end{tabular}
    \begin{tablenotes}
       \footnotesize
       \item[1] The Ratio represents the percentage of time consumed for training with a subset of features versus training with all(1083) features
     \end{tablenotes}
\end{threeparttable}
\end{center}
\end{table}

\subsection{Comparison with Related Work}
To make a comprehensive comparison between the proposed DroidRL framework and related Android malware detection methods, we conduct a comparative experiment based on the dataset in our work. The implementation of detection methods in this section refers to the code of the work \cite{Borja2023}.

Firstly, to illustrate the effect of feature selection in machine learning based Android malware detection, we compare DroidRL with other Android malware detection approaches without applying the feature selection method. 
The results are displayed in Table \ref{compare with no selection}. The proposed DroidRL framework outperforms DroidDet \cite{DroidDet}, HMMDetector \cite{HMMDetector}, and Drebin \cite{Arp2014} with higher accuracy and fewer features used in the detection, which demonstrates the features selected by reinforcement learning are highly relevant to the malware attributes. 
With 190,072 more features extracted compared to DroidRL, MamaDroid \cite{MaMaDroid} only obtains a 0.033 higher accuracy, increasing accuracy at the huge running cost. 
It is highly time-consuming to extract 190,096 features and detect malware with so many features.
Android malware detection is much more computationally efficient with valid feature subsets containing only 24 features selected by reinforcement learning used in our model.

\begin{table}[H]
\begin{center}
    \caption{\label{compare with no selection}\textbf{Comparison with detection approaches without feature selection}}
    \begin{tabular}{|l|l|l|l|}
    \hline
    Method & Number of Features & Accuracy & ML Based Detection Model\\
    \hline
    MamaDroid\cite{MaMaDroid} & 190,096 & 0.989 & Random Forest\\
    \hline
    DroidDet\cite{DroidDet} & 3,122 & 0.921 & Rotation Forest\\
    \hline
    HMMDetector\cite{HMMDetector}& - & 0.871 & Hidden Markov Model, Random Forest\\
    \hline
    Drebin\cite{Arp2014}& 545,356 & 0.881 & Support Vector Machine\\
    \hline
    \textbf{DroidRL} (ours) & \textbf{24} & \textbf{0.956} & \textbf{Random Forest}\\
    \hline
    \end{tabular}
\end{center}
\end{table}

To further demonstrate the optimality of the features selected by DroidRL, the performance of DroidRL is compared with other feature selection methods used for machine learning based Android malware detection.
The results of these experiments are shown in Table \ref{compare with selection} with the number of features and detection performance accuracy as indicators. 
We implemented these methods with a specified number of features used in detection as shown in the table (not necessarily the same number in the original work).
Compared with other traditional feature selection methods listed in the table, DroidRL obtains higher accuracy with a smaller number of features used in the detection, displaying that reinforcement learning retains its power to filter optimal features. Because, as the wrapper-based feature selection method that can utilize the feedback from the classifier to evaluate the feature subset, DroidRL can select an optimal subset of features.

\begin{table}[H]
\begin{center}
    \caption{\label{compare with selection}\textbf{Comparison with different feature selection approaches}}
    \begin{tabular}{|l|l|l|l|l|}
    \hline
    Method & Number of Features & Accuracy & Feature Selection Method & ML Based Detection Model\\
    \hline
    ICCDetector \cite{ICCDetector} & 40 & 0.948 & Correlation-based Feature Selection & Support Vector Machine\\
    \hline
    BasicBlocks \cite{BasicBlocks} & 45 & 0.947 & Information Gain & Random Forest\\
    \hline
    \textbf{DroidRL} (ours) & \textbf{24} & \textbf{0.956} & \textbf{Reinforcement Learning} & \textbf{Random Forest}\\
    \hline
    \end{tabular}
\end{center}
\end{table}

\begin{table}[H]
\begin{center}
    \caption{\label{performence1}\textbf{Detailed performance of each classifier(1)}}
    \begin{tabular}{|l|c|c|c|c|c|c|c|c|c|}
    \hline
    \diagbox{\textbf{Classifier}}{\textbf{Features Used}}&
    6& 7& 8& 9& 10& 11& 12& 13& 14\\
    \hline
    \textbf{Decision Tree}&
    \textbf{0.929}& \textbf{0.927}& 0.932& \textbf{0.943}& 0.938& 0.935& \textbf{0.932}& 0.938 &\textbf{0.942}\\
    \hline
    \textbf{Random Forest}&
     \textbf{0.929}& \textbf{0.927}& \textbf{0.933}& 0.942& \textbf{0.940}& \textbf{0.936}& \textbf{0.932}& \textbf{0.940} &\textbf{0.942}\\
    \hline
    \textbf{KNN}&
     0.911& 0.732& 0.811& 0.891& 0.810& 0.660& 0.920& 0.860 &0.820\\
    \hline
    \textbf{SVM}&
     0.928& \textbf{0.927}& 0.930& 0.941& 0.935& \textbf{0.936}& \textbf{0.932}& \textbf{0.940}& 0.938\\
     \hline
    \end{tabular}
\end{center}
\end{table}

\begin{table}[H]
\begin{center}
    \caption{\label{performence2}\textbf{Detailed performance of each classifier(2)}}
    \begin{tabular}{|l|c|c|c|c|c|c|c|c|c|c|}
    \hline
    \diagbox{\textbf{Classifier}}{\textbf{Features Used}}&
    15& 16& 17& 18& 19& 20& 21& 22& 23& 24\\
    \hline
    \textbf{Decision Tree}&
    \textbf{0.936}& 0.941& 0.946& \textbf{0.944}& 0.945& \textbf{0.943}& 0.940& 0.950 &\textbf{0.948} &0.955 \\
    \hline
    \textbf{Random Forest}&
    \textbf{0.936}& \textbf{0.942}& \textbf{0.948}& \textbf{0.944}& \textbf{0.946}& \textbf{0.943}& 0.939& \textbf{0.952} &\textbf{0.948} &\textbf{0.956} \\
    \hline
    \textbf{KNN}&
    0.793& 0.940& 0.934& 0.940& 0.936& 0.936& 0.937& 0.938& 0.945& 0.945 \\
    \hline
    \textbf{SVM}&
    \textbf{0.936}& 0.939& 0.941& 0.939& 0.942& 0.941& 0.938& 0.943 &0.945 &0.951 \\
    \hline
    \end{tabular}
\end{center}
\end{table}

\section{CONCLUSION}
The proposed DroidRL applies the DDQN algorithm to the feature selection phase to select the optimal subset of features for Android malware detection. Especially, the RNN-like network is applied as the decision network in DDQN for its capability of processing variable-length sequences. For the purpose of finding the correlation between features, DroidRL uses word embedding to semantically represent the features. During the training phase, the train-off policy is used to increase the feature search space of the DroidRL. Experiments on Drebin and Androzoo demonstrate that the DroidRL framework shows better performance than the traditional static feature extraction models, markedly improving detection performance on a variety of classifiers.
The DroidRL is shown to be effective on feature selection tasks and it will hopefully serve as an element to build up a robust malware detector in the future.

\begin{table}[H]
\begin{center}
    \caption{\label{hyperparameters}\textbf{Hyperparameter setting}}
    \begin{tabular}{|c|c|l|}
    \hline
    Parameter & Value & Description \\
    \hline
    replay start size & 50,000 &
    The number of steps carried out by the agency using uniform random policy\\
    \hline
    replay buffer size & 200,000 &
    The capacity of replay buffer memory\\
    \hline
    batch size & 32 &
    The number of training cases over which each gradient descent update is computed \\
    \hline
    discount factor& 0.99 &
    The factor that determines the importance of future rewards in the Q-learning\\
    \hline
    start learning rate& 0.0003 &
    The start factor of the linear decay scheduler\\
    \hline
    training interval& 5 &
    The frequency that the decision network is trained\\
    \hline
    \end{tabular}
\end{center}
\end{table}

\bibliographystyle{cas-model2-names}
\bibliography{refer,IEEEexample}

\newpage
\bio{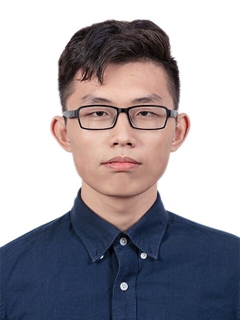}
Yinwei Wu studies in Software College of Sichuan University, Chengdu, China. He is now involved in research work on information security. His research interests include software security, deep learning and reinforcement learning.
\endbio
\vspace{55pt}

\bio{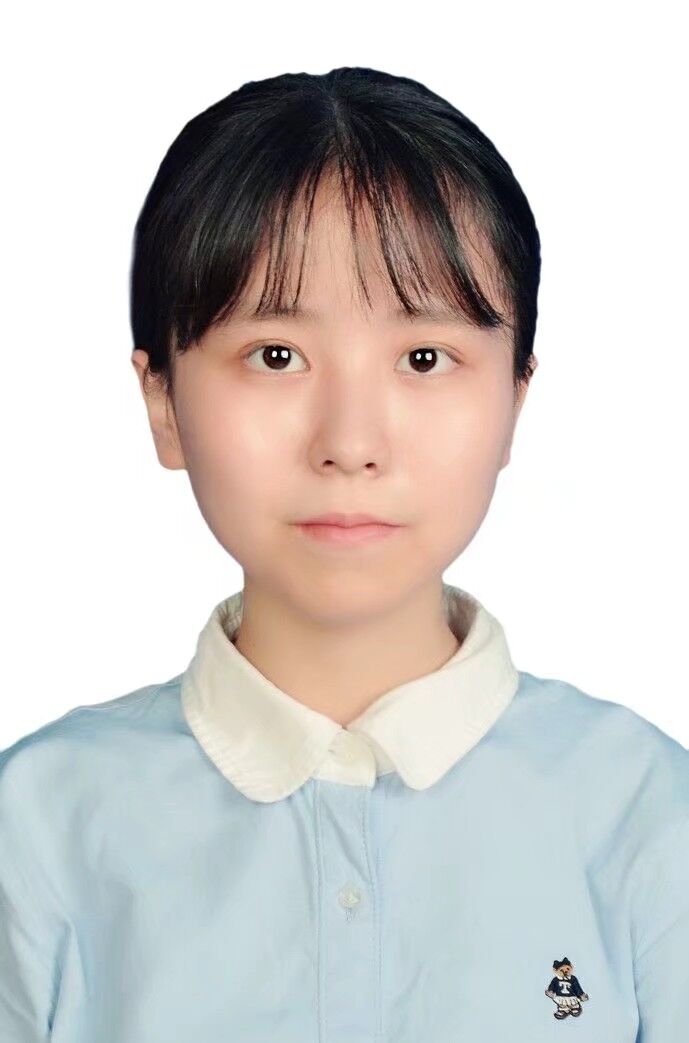}
Meijin Li is from Sichuan University, Chengdu, China. She is currently engaged in research in the field of network security and is interested in machine learning and mobile security.
\endbio
\vspace{65pt}

\bio{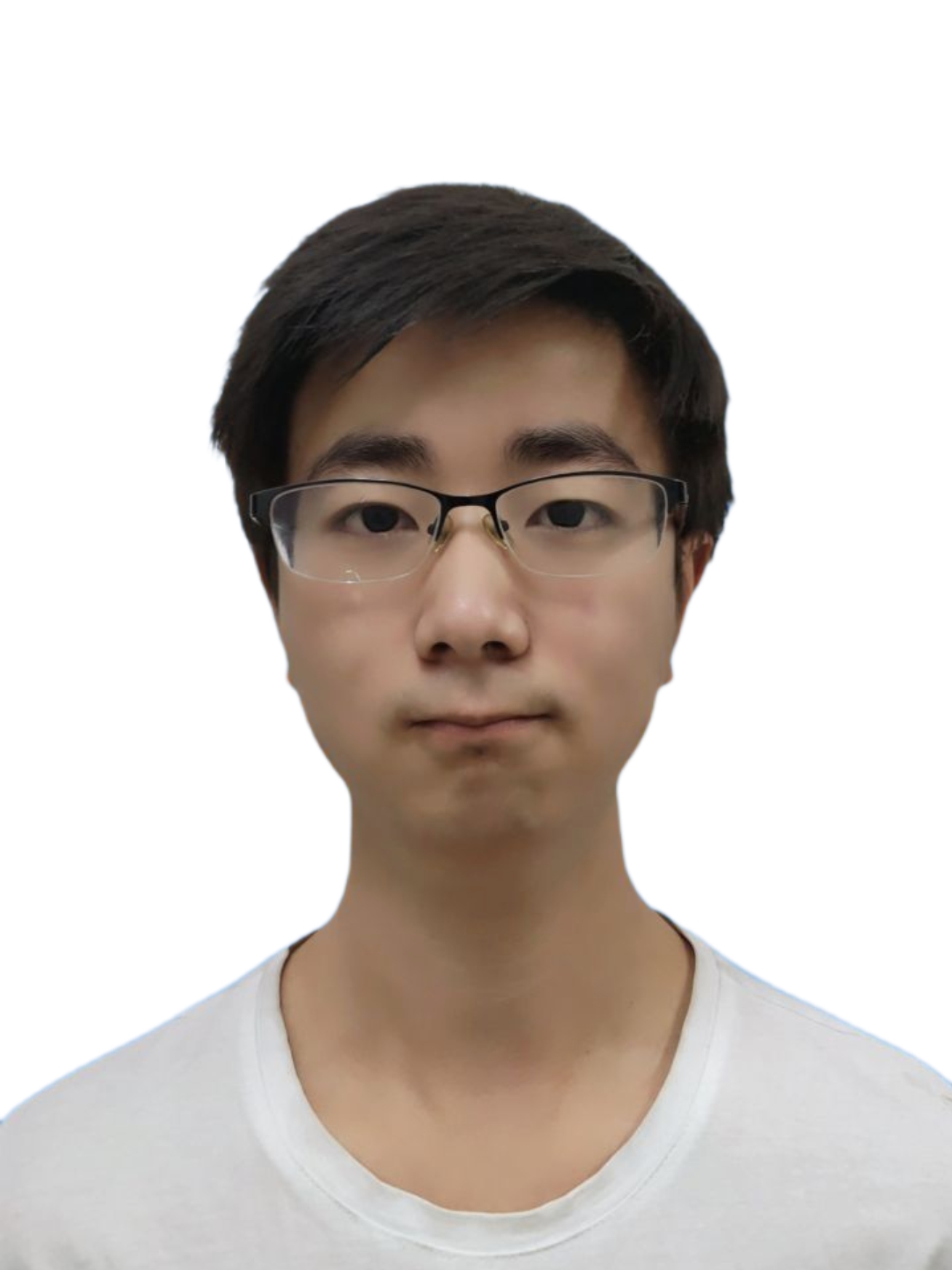}
Zeng Qi is expected to receive a bachelor's degree in Computer Science and Technology from Sichuan University, Chengdu in 2023. His research interests include machine learning and software security
\endbio
\newpage
\vspace{55pt}
\bio{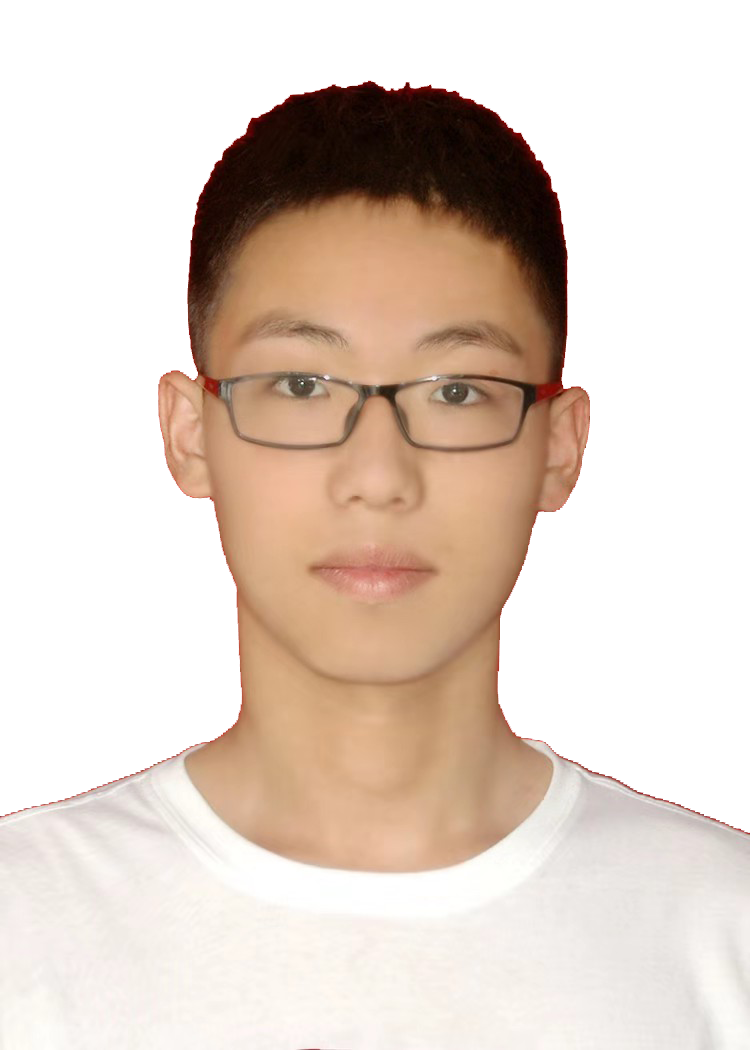}
Tao Yang is currently pursuing a bachelor's degree in Computer Science and Technology from Sichuan University, Chengdu in 2022. He is recently involved in research work on information security. His research interests include android malware detection and machine learning.
\endbio
\vspace{55pt}

\bio{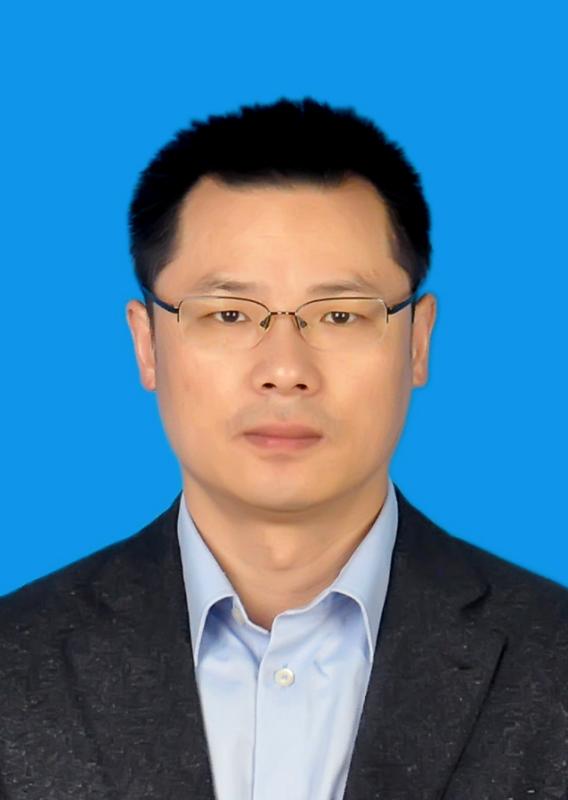}
Junfeng Wang received the M.S. degree in Computer Application Technology from Chongqing University of Posts and Telecommunications, Chongqing in 2001 and Ph.D. degree in Computer Science from University of Electronic Science and Technology of China, Chengdu in 2004. From July 2004 to August 2006, he held a postdoctoral position in Institute of Software, Chinese Academy of Sciences. From August 2006, Dr. Wang is with the College of Computer Science and the School of Aeronautics \& Astronautics, Sichuan University as a professor. He is currently serving as an associate
editor for IEEE Access, IEEE Internet of Things and Security and Communication Networks, etc. His recent research interests include network and information security, spatial information networks and data mining.
\endbio

\vspace{35pt}
\bio{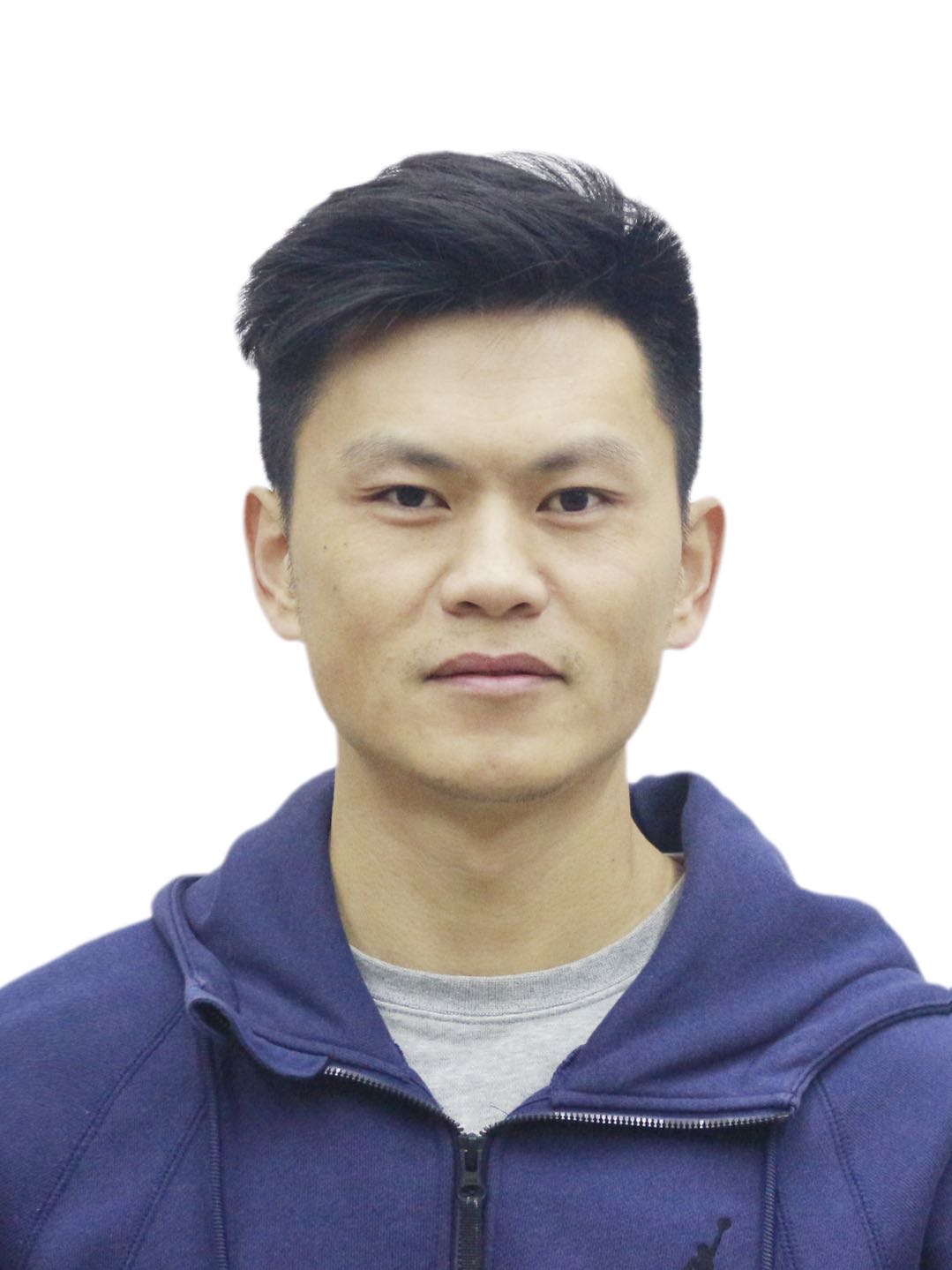}
Zhiyang Fang received his Ph.D. degree in Computer Science and Technology from Sichuan University, Chengdu in 2020. He is currently involved in research work on information security. His research interests include software security, deep learning and software engineering.
\endbio
\vspace{55pt}

\bio{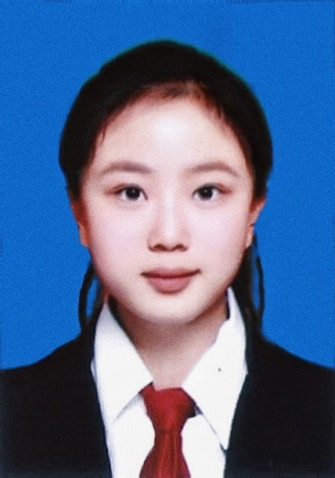}
Luyu Cheng studies in Business School from Sichuan University, Chengdu, China. Her research interests include supply chain management, data analysis, etc.
\endbio

\end{document}